\pdfoutput = 1
\documentclass{llncs}
\usepackage{makeidx}  
\usepackage{url}
\usepackage{color}
\usepackage{graphicx}
\usepackage{multirow}
\usepackage{array}
\usepackage{booktabs}
\usepackage{rotating}

\usepackage{amsmath}
\usepackage{amssymb}
\usepackage{tabularx}
\usepackage{algorithm}
\usepackage{algorithmic}
\usepackage{calrsfs}
\usepackage{appendix}
\usepackage{subfigure}
\usepackage{upgreek}
\usepackage{hyperref}
\DeclareMathAlphabet{\mathcal}{OMS}{cmsy}{m}{n}
\usepackage{color}
\usepackage{adjustbox}

\newcommand{\tabincell}[2]{\begin{tabular}{@{}#1@{}}#2\end{tabular}}

\newcommand{\abs}[1]{\left|#1\right|}
\newcommand{\norm}[1]{\left\lVert#1\right\rVert}
\newcommand{\printfnsymbol}[1]{
\textsuperscript{\@fnsymbol{#1}}
}

\begin{document}
\mainmatter                    
\title{A Multi-task Deep Feature Selection Method for Brain Imaging Genetics}
\titlerunning{A Multi-task Deep Feature Selection Method for Brain Imaging Genetics}

\author{
Chenglin Yu, Dingnan Cui, Muheng Shang, Shu Zhang, Lei Guo, Junwei Han, Lei Du\thanks{Correspondence to Lei Du (dulei@nwpu.edu.cn).This work was supported by NSFC [61973255, 61936007, 61602384]; Natural Science Basic Research Plan in Shaanxi Province of China [2020JM-142]; China Postdoctoral Science Foundation [2020T130537] at Northwestern Polytechnical University. This work was also supported by the Shanghai Municipal Science and Technology Major Project [2018SHZDZX01], LCNBI and ZJLab.}
and Alzheimer's Disease Neuroimaging Initiative\thanks{Data used in preparation of this article were obtained from the Alzheimer's Disease Neuroimaging Initiative (ADNI) database (adni.loni.usc.edu). As such, the investigators within the ADNI contributed to the design and implementation of ADNI and/or provided data but did not participate in analysis or writing of this report. A complete listing of ADNI investigators can be found at: http://adni.loni.usc.edu/wp-content/uploads/how\_to\_apply/ADNI\_Acknowledgement\_List.pdf.}
}
\authorrunning{Du et al.}
\institute{School of Automation, Northwestern Polytechnical University, Xi'an, China}

\maketitle

\begin{abstract}
Using brain imaging quantitative traits (QTs) to identify the genetic risk factors is an important research topic in imaging genetics. Many efforts have been made via building linear models, e.g. linear regression (LR), to extract the association between imaging QTs and genetic factors such as single nucleotide polymorphisms (SNPs). However, to the best of our knowledge, these linear models could not fully uncover the complicated relationship due to the loci's elusive and diverse impacts on imaging QTs. Though deep learning models can extract the nonlinear relationship, they could not select relevant genetic factors. In this paper, we proposed a novel multi-task deep feature selection (MTDFS) method for brain imaging genetics. MTDFS first adds a multi-task one-to-one layer and imposes a hybrid sparsity-inducing penalty to select relevant SNPs making significant contributions to abnormal imaging QTs. It then builds a multi-task deep neural network to model the complicated associations between imaging QTs and SNPs. MTDFS can not only extract the nonlinear relationship but also arms the deep neural network with the feature selection capability. We compared MTDFS to both LR and single-task DFS (DFS) methods on the real neuroimaging genetic data. The experimental results showed that MTDFS performed better than both LR and DFS in terms of the QT-SNP relationship identification and feature selection. In a word, MTDFS is powerful for identifying risk loci and could be a great supplement to the method library for brain imaging genetics.

\end{abstract}

\section{Introduction}

In recent years, brain imaging genetics attracts more and more attention owing to its improved power in identifying genetic risk factors than case-control studies \cite{potkin2009,shen2020}. In brain imaging genetics, the imaging quantitative traits (QTs) and single nucleotide polymorphisms (SNPs) are usually analyzed jointly, and further could help reveal novel risk loci for brain disorders such as Alzheimer's disease (AD) \cite{shen2020} and schizophrenia (SZ) \cite{arslan2018}.

Up to now, there have been many efforts made for imaging genetics. For example, based on the univariate method, Shen and Thompson \cite{shen2010ni} used the structural brain imaging measures and confirmed several risk loci for AD. Wang \emph{et al.} \cite{wang2012bioinfo1} used SNPs to predict imaging QTs based on the multi-task linear regression (LR). Sparse canonical correlation analysis (SCCA) was also utilized to study the associations between imaging QTs and SNPs, which usually combined with the feature selection techniques to identify risk loci \cite{du2020mia,fang2016joint,lin2013mia,yan2014eccb}. A common issue of these methods is that they are linear models. Therefore, they might insufficient to reveal the complicated yet challenging mechanism of the heritability of human brain that genetic factors could hardly follow a linear relationship to affect the brain structure and function, as well as brain disorders \cite{grasby2020,hibar2015}.

Recently, deep neural network (DNN) has shown great success in many applications such as image classification and objective detection \cite{krizhevsky2012,lecun2015}. The DNN can extract nonlinear relationship between imaging QTs and SNPs, but it usually suffers from the interpretation issue. In other word, a conventional DNN model cannot tell us that which SNPs, within a large candidate SNP set, contribute significantly to the imaging QTs. Therefore, it is essential to design novel DNN models with good interpretable ability, which has the potential to identify meaningful loci that linear models cannot.

In this paper, we proposed a multi-task deep neural network based feature selection (MTDFS) method to model the nonlinear correlation between imaging QTs and SNP, as well as identify relevant SNPs. First, to figure out the relevant SNPs, MTDFS introduces a sparse multi-task one-to-one layer in front of the DNN, and imposes sparsity-inducing penalties on this layer. This setup implements the feature selection in terms of SNPs, and thus makes the MTDFS interpretable. Second, MTDFS builds a multi-task DNN based prediction model where the SNPs are independent variables and imaging QTs are dependent variables. This can model the nonlinear relationship between SNPs and QTs. We used real neuroimaging genetic data downloaded from Alzheimer's Disease Neuroimaging Initiative (ADNI) database, and compared MTDFS with one multi-task linear method (LR for short) \cite{wang2012bioinfo1} and one single-task DFS (DFS for short) \cite{li2015}. The results showed that MTDFS held higher correlation coefficient (CC) and lower root mean square (RMSE) than both LR and DFS. Interestingly, MTDFS also obtained better weight profile, since most of its identified SNPs were related to AD. In contrast, LR reported too many relevant SNPs which was hard to interpret, and DFS's weights were also denser than MTDFS which means it was weaker than MTDFS. In summary, MTDFS improved the performance of both LR and DFS, showing that it could be very promising in brain imaging genetics.

\section{Methods}

\subsection{Overview}

Using brain imaging QTs as dependent variables has shown great success in identifying genetic risk loci. In generally, the brain region of interest (ROI) areas are pre-selected, and their neuroimaging measurements (QTs) are then extracted. After that, a regression or multi-task regression model is built to predict these imaging QTs using SNP data as independent variables \cite{wang2012bioinfo1}. Since not all SNPs are effective for a specific brain disorder, the regularization techniques are employed to select those SNPs of relevance. However, these methods can only find out the linear relationship. As analyzed earlier, the human genome could nonlinearly affect the brain structure and function, thereby nonlinearly influencing those abnormal imaging QTs. DNN could be a desirable alternative, but a critical issue is that it cannot select features for the input space. Li \emph{et al.} proposed a DNN with feature selection by adding a sparse one-to-one layer in front of the deep neural network \cite{li2015}. But this model only applies to single task, which indicates that it ignores the relationship among multiple interrelated tasks, i.e. predicting multiple correlated imaging QTs based on SNPs in this paper.

\begin{figure}[htbp]
  \begin{center}
    \includegraphics[width = 0.9\linewidth] {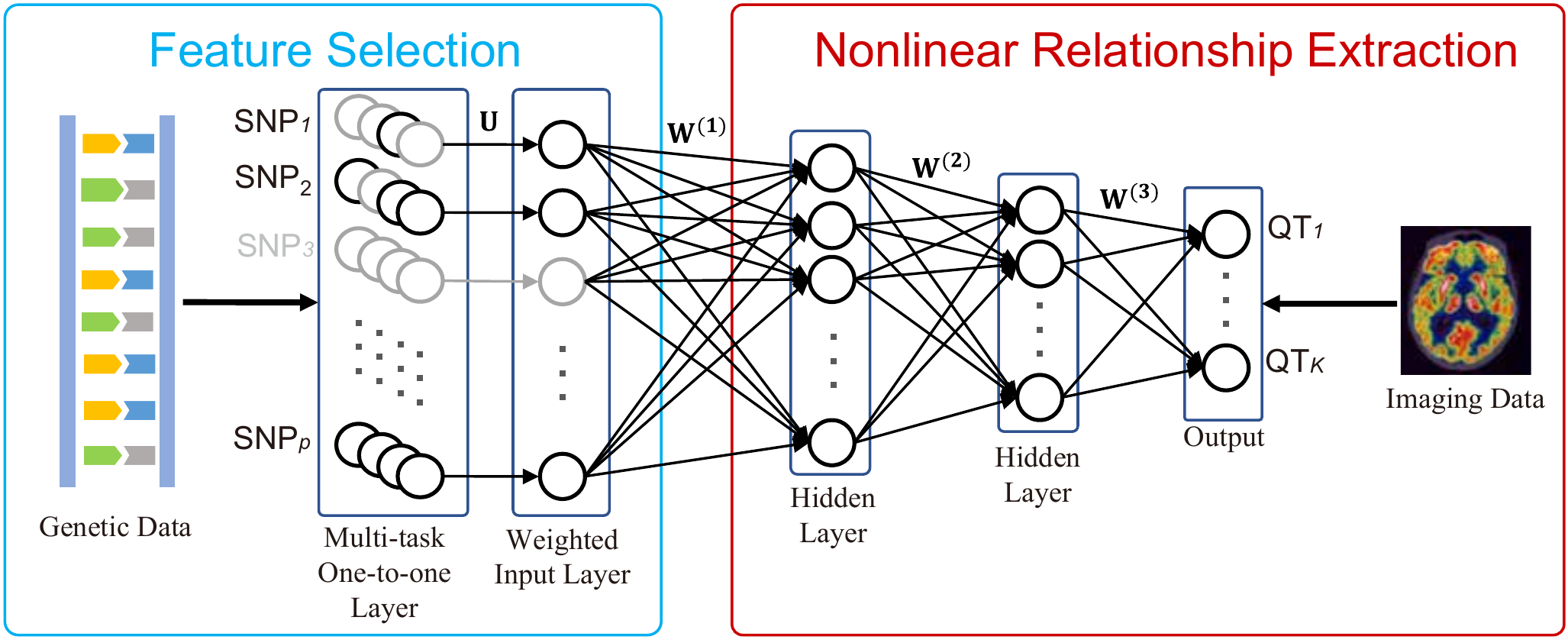}
    \caption{Overview of the proposed MTDFS model. The framework consists of two components: feature selection and nonlinear relationship extraction. The feature section component (blue block) adds a sparse multi-task one-to-one layer in front of the conventional neural network, and then imposes a hybrid penalties on this layer. On this account, only correlated SNPs will be fed into the nonlinear relationship extraction component. The nonlinear relationship extraction component (red block) first extracts the nonlinear relationship layer-by-layer, and then employs the multi-task regression to predict multiple interrelated imaging QTs. Therefore, MTDFS can not only extract nonlinear relationships, but also selects features of interest.}
    \label{fig:framework}
  \end{center}
\end{figure}

To extract the nonlinear relationship between multiple imaging QTs and SNPs, and with aim to identify relevant SNPs, we propose the multi-task deep feature selection (MTDFS) method. MTDFS adds a sparse multi-task one-to-one layer which can select those relevant SNPs, and then builds a multi-task DNN model to predict multiple interrelated imaging QTs based on SNPs. The framework of MTDFS is presented in Fig.~\ref{fig:framework}. Formally, MTDFS is defined as,
\begin{equation} \label{eq:mtdfs}
\min_{\mathbf{U},\mathbf{W}} \mathcal{L}(\mathbf{U},\mathbf{W}) + \mathcal{R}(\mathbf{U}),
\end{equation}
where $\mathcal{R}(\mathbf{U})$ is the regularization term corresponding to the feature selection component, and $\mathcal{L}(\mathbf{U},\mathbf{W})$ is the loss function for the nonlinear relationship extraction component. Next, we will introduce both components in details.

\subsection{Feature Selection}

To select a feature subset out of the whole input space, we add a sparse multi-task one-to-one layer in front of the conventional DNN. It is worthy nothing that our model has better generalization ability than DFS since MTDFS will reduce to DFS when there is only one task \cite{li2015}. As shown in Fig.~\ref{fig:framework}, the regularization techniques are used for this additional layer. Suppose there are $T$ imaging QTs (tasks), we will have $T$ one-to-one layers, with each corresponding to one imaging QT (task). To enable a reasonable model, we here are interested in three distinct types of sparsity, i.e. the element-sparsity, individual-sparsity, and group-sparsity \cite{du2020tmi}. Specifically, denoting the weight for the multi-task one-to-one layer as $\mathbf{U}$, the regularization terms are defined as follows,
\begin{equation} \label{eq:penalty}
\mathcal{R}(\mathbf{U}) = \lambda \norm{\mathbf{U}}_{\rm G_{2,1}} + \beta \norm{\mathbf{U}}_{2,1} + \gamma \norm{\mathbf{U}}_{1,1},
\end{equation}
where $\lambda$, $\beta$ and $\gamma$ are nonnegative parameters to control the layer sparsity.

The $\rm G_{2,1}$-norm indicates the group-sparsity, and can select a group of SNPs in the same linkage disequilibrium (LD) block for multiple interrelated tasks. According to \cite{wang2012bioinfo1}, {\scriptsize $\norm{\mathbf{U}}_{\rm G_{2,1}} = \sum_{m=1}^{M} \sqrt{\sum_{i \in g_m} \sum_{t=1}^{T} (u_{it})^2}$}, where $M$ is the number of LD, and $g_m$ indicates the $m$-th LD set. The $\ell_{2,1}$-norm ({\scriptsize $\norm{\mathbf{U}}_{2,1} = \sum_{i=1}^{p} \norm{\mathbf{u}^i}_2 = \sum_{i=1}^{p} \sqrt{\sum_{t=1}^{T} (u_{it})^2}$}, where $p$ is the number of SNPs) selects a relevant SNP for multiple interrelated tasks simultaneously, and thus this is individual-sparsity since it can identify a single SNP that shared among multiple tasks. Finally, $\ell_{1,1}$-norm ({\scriptsize$\norm{\mathbf{U}}_{1,1}=\sum_i\sum_j \abs{u_{ij}}$}) refers to the element-sparsity, and thus help determine whether a SNP is effective for a specific imaging QT. This hybrid penalty enables a diverse and flexible feature selection for the MTDFS model.

\subsection{Nonlinear Relationship Extraction}

The nonlinear relationship extraction component is a deep neural network. In the MTDFS model, we denote the parameters of DNN as $\mathbf{W}$, and the parameters of the $k$-th layer as $\mathbf{W}^{(k)}$. Then formally, the loss function of the MTDFS model is as follows,
\begin{equation} \label{eq:loss}
\mathcal{L}(\mathbf{U},\mathbf{W}) = F(\mathbf{U}, \mathbf{W}^{(1)}, \mathbf{W}^{(2)}, \cdots, \mathbf{W}^{(K)}),
\end{equation}
where $F(\mathbf{U}, \mathbf{W}^{(1)}, \mathbf{W}^{(2)}, \cdots, \mathbf{W}^{(K)})$ represents the DNN objective, and $K$ is the number of the layers. Since we have multiple interrelated imaging QTs corresponding to multiple tasks, these $\mathbf{U}$ and $\mathbf{W}$'s will be jointly optimized following the multi-task learning.

In this paper, imaging QTs are continuous, and thus the output layer of DNN is a regression function. Certainly, other prediction model such the negative log-likelihood (NLL) or softmax functions can also be used if applicable.

\section{Experiments}

\subsection{Real Neuroimaging Genetic Data}

The genotying and brain imaging data used in this paper were downloaded from the Alzheimer's Disease Neuroimaging Initiative (ADNI) database (adni.loni.usc.edu). One primary goal of ADNI has been to test whether serial magnetic resonance imaging (MRI), positron emission tomography (PET), other biological markers, and clinical and neuropsychological assessment can be combined to measure the progression of mild cognitive impairment (MCI) and early Alzheimer's disease (AD). For up-to-date information, see www.adni-info.org.

\begin{table}[htbp]  \scriptsize
  \centering
  \caption{Participant characteristics.}
    \begin{tabular}{rccc}
    \toprule
          & HC    & MCI   & AD \\
    \midrule
    Num   & 182   & 292   & 281 \\
    Gender (M/F, \%) & 48.90/51.10 & 48.63/51.37 & 53.38/46.62 \\
    Handedness (R/L, \%) & 89.56/10.44 & 88.70/11.30 & 90.39/9.61 \\
    Age (mean$\pm$std) & 73.93$\pm$5.51 & 70.90$\pm$6.84 & 72.61$\pm$8.15 \\
    Education (mean$\pm$std) & 16.43$\pm$2.68 & 16.18$\pm$2.68 & 15.95$\pm$2.82 \\
    \bottomrule
    \end{tabular}%
  \label{tab:demographics}%
\end{table}

The 18-F florbetapir PET (AV45) scans and genetic data were downloaded from the LONI website (adni.loni.usc.edu). There were 281 AD, 292 MCI and 182 healthy control (HC) non-Hispanic Caucasian participants, and their details were presented in Table~\ref{tab:demographics}. The PET scans were preprocessed following the pipeline including averaging, alignment to a standard space, resample to a standard image and voxel size, smoothness to a uniform resolution and normalization to a cerebellar gray matter reference region which finally yielded standardized uptake value ratio images \cite{jagust2010}. Before experiments, we normalized these images to the Montreal Neurological Institute (MNI) space as 8 m$^3$ voxels based on the MRI segmentation. We obtained ROI level amyloid measurements based on the MarsBaR AAL atlas. Since this experiment is AD-oriented, we carefully selected ten (AD-related) ROI level imaging QTs, including eight frontal areas and two olfactory areas. In addition, we utilized 2,000 SNPs from chromosome 19 nearing the AD risk genes such as \emph{APOE}. The LD block was pre-calculated. The goal was to evaluate whether the nonlinear relationship between these PET scans and SNPs was better than the linear one. In the meanwhile, we also aimed to identify the AD-risk SNPs during the nonlinear modeling.

\subsection{Implementation Details}

We compared MTDFS to two most related methods, i.e. LR \cite{wang2012bioinfo1} and DFS \cite{li2015}, to access the performance. The LR employs a multi-task learning paradigm with the same hybrid sparsity-inducing penalty. The DFS uses the single-task DNN to model the nonlinear relationship and an additional one-to-one layer to select features. Comparing to LR could evaluate the performance of the nonlinear relationship extraction component, and the performance of multi-task one-to-one layer can be evaluated by comparing to the DFS. Therefore, using both LR and DFS could fully evaluate our proposed MTDFS. Those other linear models such as sparse canonical correlation analysis \cite{du2020mia,fang2016joint,lin2013mia,yan2014eccb} were excluded since they were not designed for prediction tasks, thereby being different to MTDFS.

We here used ten imaging QTs and 2,000 SNPs, which resulted in ten interrelated tasks corresponding to ten sparse one-to-one layers. Therefore, the dimension of $\mathbf{U}$ was $2000 \times 10$, and the regularization terms was applied to $\mathbf{U}$ as defined in Eq.~(\ref{eq:penalty}). There were 2,000 units in the weighted input layer. The first hidden layer had 128 units and the second one had 64 units. Since we had ten imaging QTs, the output layer had ten units accordingly. DFS employed the same DNN structure as MTDFS. The LR model utilized the multi-task regression with the same hybrid penalty as defined in Eq.~(\ref{eq:penalty}). We used the five-fold cross-validation to tune parameters (candidate set[0.00001, 0.0001, 0.001, 0.01, 0.1]). All methods used the same setup including the data partition, number of iterations (1,000) to ensure a fair comparison.

\subsection{Improved Imaging Phenotype Prediction}

\begin{table*}[htbp] \scriptsize
  \centering
  \caption{The average RMSE along with the standard deviation (in the parentheses) of ten tasks. The best values in the testing set were shown in bold.}
  \begin{adjustbox}{max width=0.9\textwidth}
    \begin{tabular}{c|l|cccccccccc}
    \toprule
    \multicolumn{2}{c}{} & Task 1 & Task 2 & Task 3 & Task 4 & Task 5 & Task 6 & Task 7 & Task 8 & Task 9 & Task 10\\
    \midrule
    & LR & \tabincell{l}{0.1884  \\ (0.006)} &\tabincell{l}{0.1760  \\ (0.004)} &\tabincell{l}{0.2235  \\ (0.006)} &\tabincell{l}{0.2259  \\ (0.007)} &\tabincell{l}{0.1974  \\ (0.004)} &\tabincell{l}{0.1965  \\ (0.005)} &\tabincell{l}{0.1897  \\ (0.005)} &\tabincell{l}{0.1971  \\ (0.005)} &\tabincell{l}{0.1783  \\ (0.004)} &\tabincell{l}{0.1784  \\ (0.004)} \\
    \multirow{3}{*}[4pt]{\tabincell{c}{Training \\ RMSE}}& DFS & \tabincell{l}{0.2137  \\ (0.007)} &\tabincell{l}{0.1981  \\ (0.004)} &\tabincell{l}{0.2601  \\ (0.008)} &\tabincell{l}{0.2624  \\ (0.009)} &\tabincell{l}{0.2267  \\ (0.005)} &\tabincell{l}{0.2244  \\ (0.004)} &\tabincell{l}{0.2182  \\ (0.006)} &\tabincell{l}{0.2269  \\ (0.007)} &\tabincell{l}{0.2026  \\ (0.005)} &\tabincell{l}{0.2019  \\ (0.004)} \\
    & MTDFS & \tabincell{l}{0.2123  \\ (0.008)} &\tabincell{l}{0.1966  \\ (0.005)} &\tabincell{l}{0.2586  \\ (0.008)} &\tabincell{l}{0.2618  \\ (0.009)} &\tabincell{l}{0.2253  \\ (0.005)} &\tabincell{l}{0.2231  \\ (0.005)} &\tabincell{l}{0.2163  \\ (0.007)} &\tabincell{l}{0.2249  \\ (0.006)} &\tabincell{l}{0.2006  \\ (0.006)} &\tabincell{l}{0.2001  \\ (0.005)} \\
    \midrule
    & LR & \tabincell{l}{0.2231  \\ (0.030)} &\tabincell{l}{0.2046  \\ (0.021)} &\tabincell{l}{0.2779  \\ (0.035)} &\tabincell{l}{0.2779  \\ (0.038)} &\tabincell{l}{0.2372  \\ (0.023)} &\tabincell{l}{0.2343  \\ (0.023)} &\tabincell{l}{0.2266  \\ (0.026)} &\tabincell{l}{0.2359  \\ (0.028)} &\tabincell{l}{0.2109  \\ (0.022)} &\tabincell{l}{0.2096  \\ (0.021)} \\
    \multirow{3}{*}[4pt]{\tabincell{l}{Testing \\ RMSE}} & DFS & \tabincell{l}{0.2179  \\ (0.035)} &\tabincell{l}{0.2024  \\ (0.025)} &\tabincell{l}{0.2685  \\ (0.039)} &\tabincell{l}{0.2712  \\ (0.042)} &\tabincell{l}{0.2314  \\ (0.027)} &\tabincell{l}{0.2295  \\ (0.025)} &\tabincell{l}{0.2230  \\ (0.032)} &\tabincell{l}{0.2315  \\ (0.032)} &\tabincell{l}{0.2062  \\ (0.028)} &\tabincell{l}{0.2051  \\ (0.025)} \\
    & MTDFS & \tabincell{l}{\textbf{0.2155} \\ (\textbf{0.032})} &\tabincell{l}{\textbf{0.1998} \\ (\textbf{0.022})} &\tabincell{l}{\textbf{0.2659} \\ (\textbf{0.037})} &\tabincell{l}{\textbf{0.2683} \\ (\textbf{0.040})} &\tabincell{l}{\textbf{0.2295} \\ (\textbf{0.024})} &\tabincell{l}{\textbf{0.2274} \\ (\textbf{0.022})} &\tabincell{l}{\textbf{0.2210} \\ (\textbf{0.029})} &\tabincell{l}{\textbf{0.2293} \\ (\textbf{0.030})} &\tabincell{l}{\textbf{0.2038} \\ (\textbf{0.025})} &\tabincell{l}{\textbf{0.2027} \\ (\textbf{0.022})} \\
    \bottomrule
    \end{tabular}%
    \end{adjustbox}
  \label{tab:rmse}%
\end{table*}%

\begin{table*}[htbp] \scriptsize
  \centering
  \caption{The average CC along with the standard deviation (in the parentheses) of ten tasks.The best values in the testing set were shown in bold.}
  \begin{adjustbox}{max width=0.9\textwidth}
    \begin{tabular}{c|l|cccccccccc}
    \toprule
    \multicolumn{2}{c}{} & Task 1 & Task 2 & Task 3 & Task 4 & Task 5 & Task 6 & Task 7 & Task 8 & Task 9 & Task 10\\
    \midrule
    & LR & \tabincell{l}{0.5957  \\ (0.012)} &\tabincell{l}{0.5846  \\ (0.014)} &\tabincell{l}{0.6498  \\ (0.016)} &\tabincell{l}{0.6426  \\ (0.015)} &\tabincell{l}{0.6223  \\ (0.012)} &\tabincell{l}{0.6182  \\ (0.01)} &\tabincell{l}{0.6289  \\ (0.012)} &\tabincell{l}{0.6287  \\ (0.011)} &\tabincell{l}{0.5975  \\ (0.019)} &\tabincell{l}{0.5867  \\ (0.016)} \\
    \multirow{3}{*}[4pt]{\tabincell{c}{Training \\ CC}} & DFS & \tabincell{l}{0.3932  \\ (0.015)} &\tabincell{l}{0.3930  \\ (0.021)} &\tabincell{l}{0.4468  \\ (0.017)} &\tabincell{l}{0.4334  \\ (0.017)} &\tabincell{l}{0.4198  \\ (0.015)} &\tabincell{l}{0.4249  \\ (0.018)} &\tabincell{l}{0.4352  \\ (0.014)} &\tabincell{l}{0.4294  \\ (0.01)} &\tabincell{l}{0.4001  \\ (0.017)} &\tabincell{l}{0.3841  \\ (0.018)} \\
    & MTDFS & \tabincell{l}{0.3984  \\ (0.016)} &\tabincell{l}{0.3983  \\ (0.021)} &\tabincell{l}{0.4529  \\ (0.021)} &\tabincell{l}{0.4333  \\ (0.019)} &\tabincell{l}{0.4247  \\ (0.017)} &\tabincell{l}{0.4285  \\ (0.021)} &\tabincell{l}{0.4426  \\ (0.017)} &\tabincell{l}{0.4384  \\ (0.013)} &\tabincell{l}{0.4081  \\ (0.019)} &\tabincell{l}{0.3908  \\ (0.018)} \\
    \midrule
    & LR & \tabincell{l}{0.2795  \\ (0.063)} &\tabincell{l}{0.2942  \\ (0.067)} &\tabincell{l}{0.3156  \\ (0.049)} &\tabincell{l}{0.3187  \\ (0.044)} &\tabincell{l}{0.3067  \\ (0.079)} &\tabincell{l}{0.3124  \\ (0.076)} &\tabincell{l}{0.3397  \\ (0.071)} &\tabincell{l}{0.3346  \\ (0.064)} &\tabincell{l}{0.2887  \\ (0.054)} &\tabincell{l}{0.2831  \\ (0.046)} \\
    \multirow{3}{*}[4pt]{\tabincell{c}{Training \\ CC}} & DFS & \tabincell{l}{0.3208  \\ (0.069)} &\tabincell{l}{0.3234  \\ (0.076)} &\tabincell{l}{0.3636  \\ (0.067)} &\tabincell{l}{0.3451  \\ (0.054)} &\tabincell{l}{0.3517  \\ (0.062)} &\tabincell{l}{0.3530  \\ (0.071)} &\tabincell{l}{0.3700  \\ (0.078)} &\tabincell{l}{0.3652  \\ (0.058)} &\tabincell{l}{0.3395  \\ (0.063)} &\tabincell{l}{0.3278  \\ (0.056)} \\
    & MTDFS & \tabincell{l}{\textbf{0.3259} \\ (\textbf{0.076})} &\tabincell{l}{\textbf{0.3289} \\ (\textbf{0.081})} &\tabincell{l}{\textbf{0.3740} \\ (\textbf{0.070})} &\tabincell{l}{\textbf{0.3562} \\ (\textbf{0.053})} &\tabincell{l}{\textbf{0.3549} \\ (\textbf{0.062})} &\tabincell{l}{\textbf{0.3581} \\ (\textbf{0.071})} &\tabincell{l}{\textbf{0.3704} \\ (\textbf{0.078})} &\tabincell{l}{\textbf{0.3694} \\ (\textbf{0.057})} &\tabincell{l}{\textbf{0.3403} \\ (\textbf{0.069})} &\tabincell{l}{\textbf{0.3289} \\ (\textbf{0.060})} \\
    \bottomrule
    \end{tabular}
    \end{adjustbox}
  \label{tab:cc}%
\end{table*}%

Two evaluation criteria were used to access the performance of three methods. Since neuroimaging QTs were continuous, we first employed the popular root mean square error (RMSE), which is the smaller the better. Besides, we utilized the correlation coefficient (CC) as another metric which was widely used too. A higher CC strands for a better performance. We presented both RMSEs and CCs in Table~\ref{tab:rmse} and Table~\ref{tab:cc}. In both tables, we can clearly observe that our MTDFS obtained better scores than both LR and DFS. This indicated that MTDFS not only predicted the dependent imaging QTs with the smallest error, but also extracted higher relationship between imaging QTs and SNPs. This revealed that DNN coupled with multi-task feature selection could obtain improved imaging phenotype prediction, demonstrating the success of MTDFS.

\subsection{Genetic Marker Selection}
\begin{figure}[htbp]
  \begin{center}
    \includegraphics[width = 0.75\linewidth] {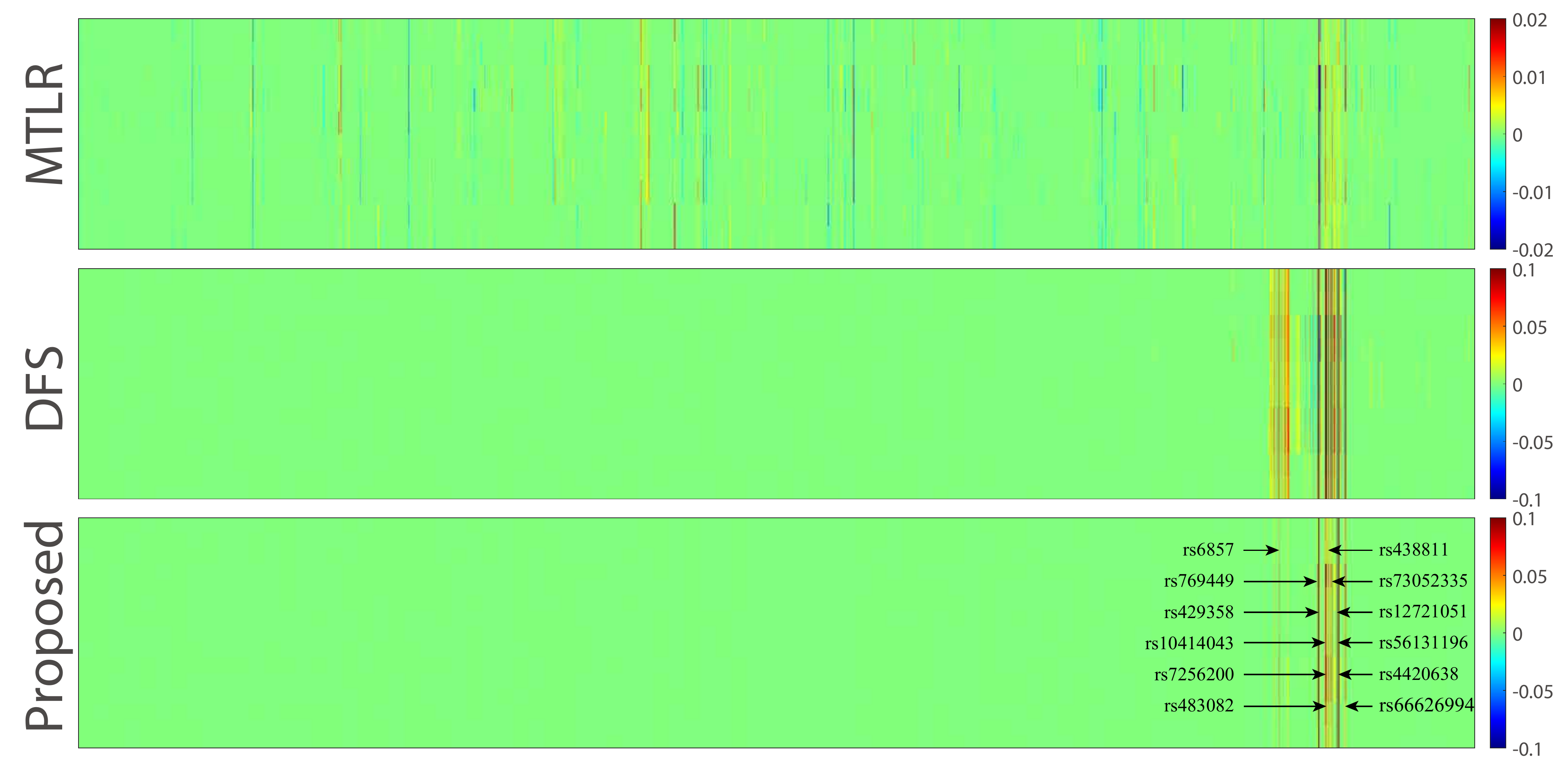}
    \caption{Comparison of regression weights. Each row corresponds to a method, and there are tens weight vectors corresponding to ten tasks for each method.}
    \label{fig:real_weight}
  \end{center}
\end{figure}

It is not surprising that DNN holds higher correlations than linear models owing to its nonlinear modeling. Therefore, it is essential to compare the feature selection results. We presented the heat map which exhibited the feature selection in Fig.~\ref{fig:real_weight}. The most relevant features were highlighted in this figure. It was clear that MTLR reported too many relevant SNPs. This was hard to interpret since identifying too many markers provides little to no useful information. DFS alleviated the drawback of MTLR, but it still identified too many markers than our method. MTDFS successfully identified a small subset out of the whole SNP candidate set. More importantly, the top identified SNPs of MTDFS, including rs429358 (\emph{APOE}), rs4420638 (\emph{APOC1}), rs12721051 (\emph{APOC1}), rs56131196 (\emph{APOC1}), rs769449 (\emph{APOE}),rs7256200 (\emph{APOC1}), rs483082 (\emph{APOC1}),rs438811 (\emph{APOC1}),rs73052335 (\emph{APOC1}) and so forth, were all correlated to AD. In addition, owing to the hybrid penalty, MTDFS also identified group structures, e.g. SNPs of the \emph{APOE} and those of the \emph{APOC1}. These results demonstrated the success of our multi-task one-to-one layer, indicating that this strategy can endow a meaningful feature selection capability to the deep neural network.

\subsection{Imaging Genetic Correlation Interpretation}

\begin{figure}[htbp]
  \begin{center}
    \includegraphics[width = 0.7\linewidth] {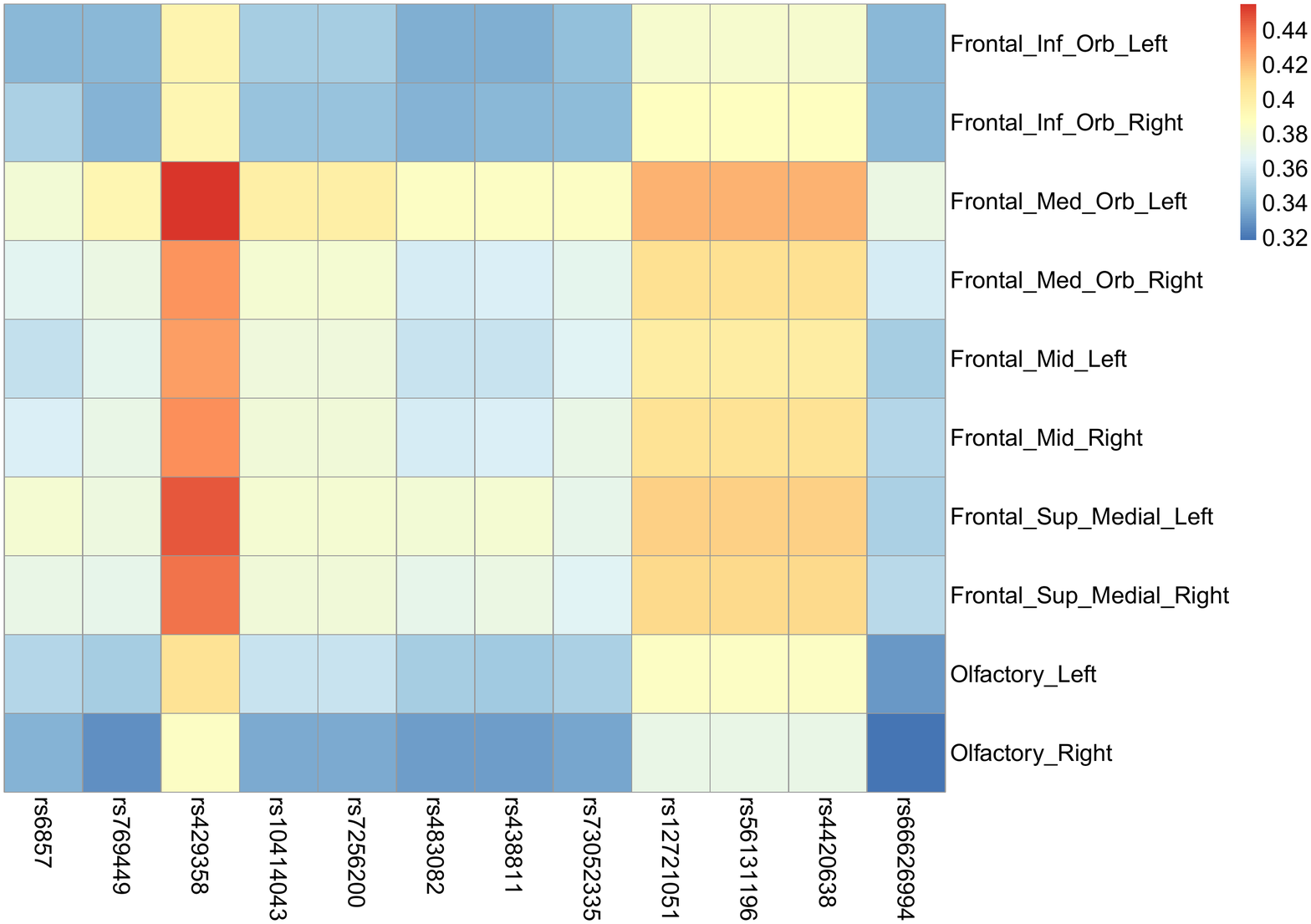}
    \caption{The pairwise correlation between imaging QTs and top selected SNPs.}
    \label{fig:img_snp}
  \end{center}
\end{figure}

To better understand the identified associations, in Fig.~\ref{fig:img_snp}, we presented the pairwise correlation between ten imaging QTs and top twelve SNPs shared by all tasks, and the Analysis of Variance (ANOVA) analysis showed that all values were significant. We observed that rs429358 had the highest weight values for all tasks, showing its importance in predicting AD-altered brain areas. In addition, rs12721051, rs56131196, and rs4420638 have the same value, which shows that they have the same importance for AD prediction. rs10414043 and rs7256200 also have the same properties.

\section{Conclusions}

Extracting the relationship between brain neuroimaging data and genetic data, as well as select relevant genetic factors, is important for brain imaging genetics. The linear model has been extensively studied but is limited since the human genome could nonlinearly affect the brain structure and function. To overcome this drawback, we proposed a multi-task deep feature selection method. MTDFS conducted nonlinear relationship extraction and feature selection simultaneously. We introduced a hybrid penalty to select features at the element-sparsity, individual-sparsity, and group-sparsity levels. Results on real neuroimaging genetic data showed that MTDFS was the most powerful approach among multi-task linear regression and single-task deep feature selection. In the future, we intend to apply the convolution network into our model since it could better identify LD for SNPs, which has the potential to better understand human brain.

\bibliographystyle{splncs04}
\bibliography{mtdfs}

\end{document}